\documentstyle[sprocl,epsf]{article}

\def\be{\begin{equation}}
\def\ee{\end{equation}}
\def\bea{\begin{eqnarray}}
\def\eea{\end{eqnarray}}
\begin{document}

\begin{flushright}
e-print JLAB-THY-97-24 \\
JINR-E2-97-202 \\ 
June 1997 \\
%hep-ph/
\end{flushright}
\vspace{1cm}

\title{THE ASYMPTOTICS OF THE TRANSITION FORM FACTOR
$\gamma\gamma^* \to \pi^o$ AND QCD SUM RULES
\footnote{Talk presented by R.~Ruskov at the Photon'97 Conference,
Egmond aan Zee, The Netherlands, May 10-15, 1997}}

\author{ A.V. RADYUSHKIN$^{* \dagger}$}

\address{$^{*}$Old Dominion University, Norfolk, VA 23529, USA;\\
Jefferson Lab, Newport News, VA 23606, USA}

\author{ R. RUSKOV$^{\dagger}$ }

\address{$^{\dagger}$Laboratory of Theoretical 
Physics, JINR, Dubna 141980, Russia}


\maketitle\abstracts{In this paper we present 
the result 
\cite{pl,rr} 
of a direct QCD  sum rule calculation of the
transition form factor  $\gamma\gamma^* \to \pi^0$
in the  region of moderately
large invariant momentum $Q^2 \ge 1 \, GeV^2$ 
 of the virtual photon.
In contrast to pQCD,  we make no assumptions
about the shape of the pion distribution amplitude $\varphi_{\pi}(x)$.
Our results  agree with the Brodsky-Lepage
proposal that the $Q^2$-dependence of this form factor
is given by an  interpolation
between its  $Q^2=0$ value fixed by the axial anomaly
and  $1/Q^2$ pQCD behaviour  for large $Q^2$,
with normalization corresponding to the asymptotic
form $\varphi_{\pi}^{as} (x)=6 f_{\pi} x (1-x)$ of
the pion distribution amplitude. Our prediction for the from factor
$F_{\gamma^*\gamma^*\pi^\circ}(q_1^2=0,q_2^2=-Q^2)$ is in good agreement
with new CLEO data.
}

 The transition $\gamma^* \gamma^* \to \pi^0$
of two virtual photons $\gamma^*$ into a neutral pion
provides an exceptional opportunity to test
QCD predictions for exclusive processes.
In the lowest order of
perturbative QCD, its asymptotic behaviour
is due to the subprocess
$\gamma^*(q_1) + \gamma^*(q_2) \to \bar q(\bar xp) + q (xp) $
with $x$ ($\bar x$) being  the fraction of the pion momentum $p$ carried
by the quark produced at the $q_1$ ($q_2)$ photon vertex.
The  relevant diagram resembles
the handbag diagram of 
DIS,
with the main difference that one should use
 the pion distribution amplitude (DA) $\varphi_{\pi}(x)$
instead of parton densities.
This  gives  good reasons to  expect that
pQCD for this process may
work at  accessible values of spacelike
photon virtualities.
The asymptotic pQCD prediction is given by \cite{bl80}\ ($\bar{x}=1-x$):
\begin{equation}
F_{\gamma^* \gamma^*  \pi^0 }^{as}(q^2, Q^2) = \frac{4\pi}{3}
\int_0^1 {{\varphi_{\pi}(x)}\over{xQ^2+\bar x q^2}} \, dx
\stackrel{q^2=0}{\longrightarrow}
\frac{4\pi}{3}
\int_0^1 {{\varphi_{\pi}(x)}\over{xQ^2}} \, dx
\equiv \frac{4\pi f_{\pi}}{3Q^2} I  .
\label{eq:gg*pipqcd}
\end{equation}
%($Q^2   \equiv - q_2^2$, $q^2 \equiv -q_1^2$, $\bar x = 1-x$).
Experimentally,
the most important situation is when one of the photons
is almost real $q^2 \approx 0$ \cite{CELLO,CLEOnew}.
In this  case,  necessary nonperturbative information
is accumulated in
the same integral $I$  (see eq.(\ref{eq:gg*pipqcd}))
that appears  in the one-gluon-exchange
diagram  for the  pion electromagnetic  form factor
\cite{pl80,blpi79,cz82}.

The  value of $I$ is sensitive to the shape of the
pion DA $\varphi_{\pi}(x)$, mainly to its end-point
behaviour. In particular,  using  the
asymptotic form
$
\varphi_{\pi}^{as}(x) = 6 f_{\pi} x \bar x
$
\cite{pl80,blpi79} gives $F_{\gamma \gamma^*  \pi^0 }^{as}(Q^2) =
4 \pi f_{\pi}/Q^2 $ for  the asymptotic
behaviour \cite{bl80}. If one takes the
Chernyak-Zhitnitsky form\cite{cz82}
$\varphi_{\pi}^{CZ}(x) = 30 f_{\pi} x\bar{x}(1-2x)^2$,
the integral $I$ increases by a sizable factor of 5/3,
and this difference can be used for experimental
discrimination between the two forms. One-loop radiative QCD corrections
to eq.(\ref{eq:gg*pipqcd}) are known \cite{braaten,kmr} and they are
under control.

For lower $Q^2$,  power corrections
 become very important.
Indeed, the asymptotic $1/Q^2$-behaviour
cannot be true in the  low-$Q^2$
region,  since  the $Q^2=0$ limit of
$F_{\gamma \gamma^*  \pi^0 }(Q^2)$
is known to be finite and
normalized by the $\pi^0 \to \gamma \gamma$ decay rate.
Theoretically\cite{ABJ},
$
F_{\gamma \gamma^*  \pi^0 }(0) =1/ \pi f_{\pi} .
$
It is natural to expect that the leading term is close to
a simple interpolation
$
\pi f_{\pi} F^{LO}_{\gamma \gamma^*  \pi^0 }(Q^2) =
1/(1+ Q^2/4 \pi^2 f_{\pi}^2)
$\cite{bl80}  
%\cite{blin}
between
the $Q^2=0$ value and the large-$Q^2$
asymptotics.
This interpolation agrees  with experiment\cite{CELLO,CLEOnew}
and implies the asymptotic form\cite{pl80,blpi79} 
of the DA for accessible  $Q^2$. It  introduces a  mass scale
$s^{\pi}_o \equiv 4 \pi^2 f_{\pi}^2 \approx 0.67~GeV^2$
close to $m_{\rho}^2$.

Consider a
three-point correlation function\footnote{Actually, it
is a common starting point both for pQCD and QCD SR approaches.}
\begin{equation}
{\cal F}_{\alpha\mu \nu}(q_1,q_2)= 2 \pi i
\int
\langle 0 |T\left\{j_{\alpha}^5(Y) J_{\mu }(X)\,J_{\nu}(0)\right\}| 0 \rangle
e^{-iq_{1}X}\,e^{ipY}  d^4X\,d^4Y \,  ,
\label{eq:corr}
\end{equation}
where $J_{\mu}$ is the EM current and the axial-vector
current has a non-zero
projection onto the neutral pion state.
The amplitude ${\cal{F}}_{\alpha\mu \nu}(q_1,q_2)$
has a pole for $p^2=m_{\pi}^2$ with residue proportional to
the form factor of interest.
The higher states  include $A_1$ and higher  broad
pseudovector resonances. Due to asymptotic freedom,
their sum for large $s$ rapidly approaches the
pQCD  spectral density ${\rho}^{PT}(s,q^2,Q^2)$.
Hence, the spectral density
of the  dispersion relation
for the relevant invariant amplitude ${\cal{F}}\left(p^2,q^2,Q^2\right)$
 can be written as
$\rho \left(s,q^2,Q^2\right)=
\pi f_{\pi}\delta(s-m_\pi^2)
\mbox{$F_{\gamma^*\gamma^*\pi^\circ}$}\left(q^2,Q^2\right)
+  \theta(s-s_o){\rho}^{PT}(s,q^2,Q^2),
$
with  the parameter $s_o$ being
the effective threshold for   higher states.
To construct   a QCD sum rule, we  calculate
the three-point function ${\cal F}(p^2,q^2,Q^2)$
and then its SVZ-transform   $\Phi (M^2,q^2,Q^2)$
as a  power  expansion in $1/M^2$  for large $M^2$.
%However, a particular   form of the
%$(1/p^2)^N$-expansion for  ${\cal F}(p^2,q^2,Q^2)$  depends
%on the   values of the photon virtualities
%$q^2 $ and  $Q^2$.

The simplest    case  is when the smaller
virtuality $q^2$  is  large:
$q^2,Q^2,-p^2 \ge 1~GeV^2$.
 Then, to produce a
  contribution with   a power behaviour $(1/p^2)^N$,
 all three currents  should be kept
close to each other: all the
  intervals $ X^2$, $ Y^2$, $(X-Y)^2$ should be small.
Taking into account the perturbative contribution and
the condensate corrections,
we obtain a  QCD sum rule\cite{pl,rr}.
For $Q^2,q^2 \gg s_0$,
keeping only the leading  $O(1/Q^2, 1/q^2)$-
%and $O(1/q^2)$ 
terms we obtain:
%the sum rule
\begin{eqnarray}
&&F_{\gamma^*\gamma^*\pi^\circ}^{LO}(q^2,Q^2)
=\frac{4\pi}{3f_{\pi}}
   \int_0^1 \frac{dx}{ ( xQ^2 + \bar x q^2)} \,
\left \{ \frac{3M^2}{2\pi^2}(1-e^{-s_0/M^2}) x\bar{x}
\right. \nonumber\\
&&+ \frac{1}{24M^2}
\langle \frac{\alpha_s}{\pi}GG\rangle [\delta(x) + \delta (\bar{x})]
\nonumber\\
&&+\left. \frac{8}{81M^4}\pi\alpha_s{\langle \bar{q}q\rangle}^2
 \biggl ( 11[\delta(x) + \delta (\bar{x})] +
2[\delta^{\prime}(x) + \delta ^{\prime}(\bar{x})]
\biggr ) \right \}
\label{eq:SRlargeQ2wf}.
\end{eqnarray}
Note, that the expression in  curly brackets
coincides with the QCD sum rule for
the pion DA
$f_{\pi} \varphi_{\pi}(x)$ (see, $e.g.,$ ref.\cite{mr}).
Hence,
the QCD sum rules approach is capable to
reproduce  the pQCD result (\ref{eq:gg*pipqcd}).

An attempt to  get
a QCD sum rule for the integral $I$ by taking  $q^2=0$
in eq.(\ref{eq:SRlargeQ2wf}) is ruined by
 power singularities
$1/q^2$, $1/q^4$
in the condensate terms.
The perturbative term in the small-$q^2$ region has
logarithms $\log q^2$
which  are a typical example
of mass singularities (see, $e.g.,$ \cite{georgietal}).  %,sterman}).
All these infrared sensitive terms are produced 
in a regime when the hard momentum flow
bypasses the soft photon vertex, $i.e.$,
%In the coordinate representation, this regime
%corresponds to keeping
the EM current $J_{\mu }(X)$ of
 the low-virtuality photon is
far away from the two other  currents $J(0),j^5(Y)$. 

Observe also, that power singularities emerge
  precisely   by  the  same
$\delta(x)$ and $\delta'(x)$ terms in eq.(\ref{eq:SRlargeQ2wf})
which  generate the two-hump form
for $\varphi_{\pi}(x)$ in the CZ-approach \cite{cz82}.
 As shown in ref.\cite{mr},
the  $\delta^{(n)}(x)$ terms result from the Taylor expansion of
nonlocal condensates like $\langle \bar q(0) q(Z) \rangle$.

Our strategy  is to subtract all these singularities from
the coefficient functions of the original OPE for the 3-point
correlation function eq.(\ref{eq:corr}). They are absorbed in this
approach by {\it universal} bilocal correlators (see refs.\cite{pl,rr}),
which can be also interpreted
as moments of the DAs  for  (almost) real photon
$
\int_0^1 y^n \phi_{\gamma}^{(i)} (y,q^2) \sim
\Pi^{(i)}_n(q^2) = \int e^{iq_1 X} \langle 0| T \{ J_{\mu }(X)
{\cal O}_n^{(i)}  (0) \}| 0 \rangle d^4X ,
$
where ${\cal O}_n^{(i)} (0)$ are operators of leading and next-to-leading
twist with $n$ covariant derivatives \cite{pl,rr}.
The bilocal contribution to the 3-point function eq.(\ref{eq:corr})
can be written in a ``parton'' form as a convolution of the 
photon DAs and some coefficient functions. The last originate from
a light cone OPE for the product  $T\{ J(0) j^5(Y) \}$.
The
amplitude ${\cal F}$   is now  a  sum of its
purely  short-distance ($SD$) (regular for $q^2=0$) and bilocal ($B$) parts.
Getting the $q^2 \to 0$ limit of $\Pi_n^{(i)}(q_1)$
requires a nonperturbative input.

After all modifications described above are made, we can write 
the QCD sum rule for the $\gamma \gamma^* \to \pi^0$ form factor
in the $q^2=0$ limit:
\begin{eqnarray}
&\,& \pi f_{\pi} F_{\gamma \gamma^* \pi^0}(Q^2) =
\int_0^{s_0}
\left \{ %1
1 - 2 \frac{Q^2-2s}{(s+Q^2)^2}
\left (s_{\rho} - \frac{s_{\rho}^2}{2 m_{\rho}^2} \right )
\right.  \nonumber \\ &+& \left.
  2\frac{Q^4-6sQ^2+3s^2}{(s+Q^2)^4} \left (\frac{s_{\rho}^2}{2}
 - \frac{s_{\rho}^3}{3  m_{\rho}^2} \right )
\right \} %1
 e^{-s/M^2}
\frac{Q^2 ds }{(s+Q^2)^2}
 \nonumber \\
&+&\frac{\pi^2}{9}
{\langle \frac{\alpha_s}{\pi}GG \rangle}
\left \{ %2
\frac{1}{2 Q^2 M^2} + \frac{1}{Q^4}
- 2 \int_0^{s_0} e^{-s/M^2} \frac{ds }{(s+Q^2)^3}
\right \} %2
 \nonumber \\ &+&
 \frac{64}{27}\pi^3\alpha_s{\langle \bar{q}q\rangle}^2
\lim_{\lambda^2 \to 0}
\left \{ %3
\frac1{2Q^2 M^4}
+ \frac{12}{Q^4 m_{\rho}^2 }
\left [ %4
\log \frac{Q^2}{\lambda ^2} -2
\right.  \right. \nonumber \\ &+& \left.  \left.
 \int_0^{s_0} e^{-s/M^2}
\left ( %5
\frac{s^2+3sQ^2+4Q^4} {(s+Q^2)^3} - \frac1{s+\lambda ^2}
\right) ds %5
\right] %4
\right. %3
\nonumber \\
&-&
\left.  %3
\frac4{Q^6}
\left [ %6
\log \frac{Q^2}{\lambda^2} -3+
\int_0^{s_0} e^{-s/M^2}
\left (  %7
\frac{s^2+3sQ^2+6Q^4} {(s+Q^2)^3} - \frac1{s+\lambda ^2}
\right) ds %7
\right] %6
\right \} .%3
\label{eq:finsr}
\end{eqnarray}
Here we model the bilocal contributions using the asymptotic form for
the DAs of the $\rho$-meson and making them  approximately
dual to the corresponding pt-contribution.
\begin{figure}[htb]
\mbox{
   \epsfxsize=8.3cm
 \epsffile{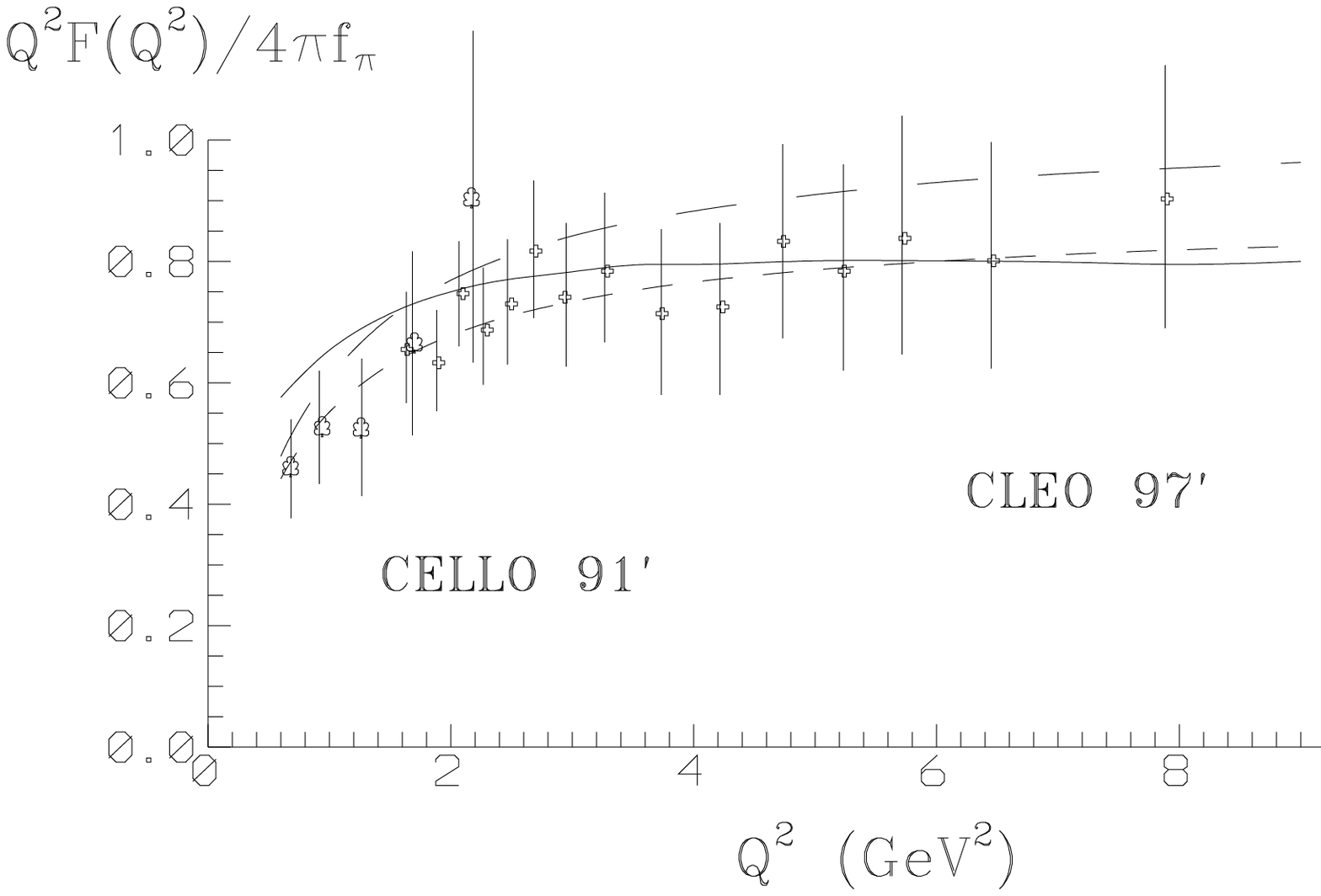}  }
  \vspace{-5.8cm}
{\caption{\label{fig:3} }}
\end{figure}
%\vspace{-0.2cm}
We use the standard values for the condensates
 and the $\rho$-meson duality interval
 $s_{\rho}=1.5~GeV^2$  \cite{svz}.
Explicit fitting procedure in
(\ref{eq:finsr}) favours the value
$s_0 \approx 0.7 \, GeV^2$ for the effective threshold \cite{pl,rr}. 
Hence, our calculations support
the local duality prescription \cite{radacta95}.

%\noindent
In Fig.\ref{fig:3}, 
we present our  curve (solid line) for
$Q^2F_{\gamma \gamma^* \pi^0}(Q^2)/4\pi f_{\pi}$
calculated from eq.(\ref{eq:finsr}) for $s_0 = 0.7 \, GeV^2$.
%and $M^2 = 0.8\, GeV^2$. 
One can observe very good agreement with
the new CLEO data \cite{CLEOnew}.
It is rather close to the
Brodsky-Lepage interpolation
formula \footnote{In fact, such
interpolation follows from the local duality 
considerations \cite{radacta95}} (long-dashed line)
and  the $\rho$-pole  approximation (short-dashed line)
$\pi f_{\pi} F^{VMD}(Q^2) = 1/(1+Q^2/m_{\rho}^2)$.
It should be noted that
the  $Q^2$-dependence of the $\rho$-pole type  emerges
due to the fact that the pion
duality interval $s_0 \approx 0.7 \, GeV^2$
is numerically  close to $m_{\rho}^2\approx 0.6\,GeV^2$.
In the region  $Q^2 > Q^2_{*} \sim 3 \, GeV^2$, our curve for
$Q^2F_{\gamma \gamma^* \pi^0}(Q^2)$
 is practically  constant, supporting
 the pQCD expectation (\ref{eq:gg*pipqcd}).  The
absolute magnitude  of our prediction  gives
  $I \approx 2.4$ for the $I$-integral with an
accuracy of about $20\%$.

Comparing the value  $I=2.4$ with
 $I^{as}=3$ and $I^{CZ}=5$, we
conclude  that our result favours
a  pion  DA
which is narrower than the asymptotic form.
Parametrizing the width of $\varphi_{\pi}(x)$  by
 a simple model $\varphi_{\pi}(x) \sim [x(1-x)]^n$,
we get that  $I=2.4$
corresponds to $n=2.5$.
The second moment  
$ \langle \xi^2\rangle \equiv \langle (x - \bar x)^ 2\rangle$
for such a function  
is 0.125
(recall that $\langle \xi^2\rangle^{as}=0.2$
while $\langle \xi^2\rangle^{CZ}=0.43$)  which agrees  with
the lattice calculation \cite{lattice}.

%\section*{Acknowledgments}
We are
grateful to S.J.Brodsky, H.G.Dosch, A.V.Efremov, O.Nachtmann,
D.J.Miller and V.Savinov for useful  discussions and comments.
The work  of AR  was supported
by the US Department of Energy under contract DE-AC05-84ER40150;
the work of RR was  supported
by Russian Foundation for Fundamental
Research, Grant $N^o$ 96-02-17631.

\end{document}